\documentclass{ws-ijgmmp}
\usepackage{graphicx}
\def\p{\partial}
\def\({\left(}
\def\){\right  )}
\def\dsum{\displaystyle\sum}
\begin{document}

\markboth{S Jamal}
{APPROXIMATE SINGULAR LAGRANGIAN}

\catchline{}{}{}{}{}

\title{A STUDY OF THE APPROXIMATE SINGULAR LAGRANGIAN - CONDITIONAL NOETHER SYMMETRIES AND FIRST INTEGRALS
}

\author{\footnotesize SAMEERAH JAMAL}

\address{School of Mathematics,
University of the Witwatersrand, Johannesburg, South Africa.
Sameerah.Jamal@wits.ac.za}

\maketitle

\begin{history}
\received{(Day Month Year)}
\revised{(Day Month Year)}
\end{history}

\begin{abstract}
The investigation of approximate symmetries of reparametrization invariant Lagrangians
of $n+1$ degrees of freedom and quadratic  velocities is presented.  We show that  extra conditions  emerge 
which give rise to approximate and conditional Noether symmetries of such constrained actions. The
Noether symmetries are the simultaneous conformal Killing vectors
 of both the kinetic metric and the potential.  In order to recover these conditional symmetry generators which would otherwise be lost in gauge fixing the lapse function entering the perturbative Lagrangian, one must
consider the lapse among the degrees of freedom. We establish a geometric framework in full generality to determine the admitted
Noether symmetries. Additionally, we obtain the corresponding first integrals (modulo a constraint equation).
For completeness, we present a pedagogical application of our method.
\end{abstract}

\ccode{PACS Nos.: 04.20.Fy; 02.20.Sv; 02.40.Ky.}

\section{Introduction}
The Action Principle
\begin{eqnarray}
S&=&\int L\left( t,x,\dot{x},N \right)~dt\nonumber\\
&=&\int\(\frac{1}{2N}g_{ij}\left( x\right)\dot{x}^{i}\dot{x%
}^{j}-NV\left( x\right)\)~dt,\quad \det g_{ij}\ne0,~ i,j=1,\ldots, n,  \label{AN}
\end{eqnarray}%
 represents a singular system consisting of $n+1$ degrees of
freedom and is quadratic in the velocities. Due to the reparametrization invariance of this action, the system of Euler-Lagrange equations
\begin{eqnarray}
{\cal H} &\dot{=}&\frac{1}{2N^2}g_{ij}\dot{x}^{i}\dot{x%
}^{j}+V=0,  \label{ANS.05} \\
{\cal E}^k &\dot{=}&\ddot{x}^k+\Gamma^k_{ij}\dot{x}^{i}\dot{x}^{j}-\frac{\dot{N}}{N}\dot{x}^{k}+N^2g^{ik}V_{,i}=0 ,  \label{ANS.06}
\end{eqnarray}
which describe its dynamics is singular. The lapse function
$N = N (t)$ is a singular degree of freedom with Euler-Lagrange equation (\ref{ANS.05}) also written as \begin{equation}\frac{\p L}{\p N}=0.\label{b1}\end{equation}

These Lagrangians are encountered in various
cosmological models -  particularly minisuperspace models where the $x^i$ represents the scale factor components and/or
possible matter fields. Subsequently new dynamical systems of such Lagrangians which
admit symmetries were found in other related areas
such as general relativity and classical mechanics.

Per contra, in much of the past and present literature, the  singular Lagrangian of the type (\ref{AN}) was relegated to that of a regular Lagrangian by setting the lapse function to $N=1$, or fixing it to some other convenient value. Under this assumption of pseudo-regularity, Christodoulakis et al.  \cite{p1} proved how this leads to the loss of a class of   symmetries called conditional symmetries \cite{p4}. Furthermore, they found that the
presence of the lapse function affects the corresponding reparametrisation
generator, i.e. the quadratic constraint which finds importance in solving
the classical equations of motion.

To encompass the notion of no regularity, instead of gauge fixing the lapse function
one considers
 the lapse as an independent degree of freedom, thereby
stipulating that (\ref{b1}) is a  constraint equation. Intrinsically, this
influences the space
of dependent variables while simultaneously placing additional restrictions on the symmetry determining conditions which will now involve $N$ and its derivatives. 

In a previous paper \cite{jgp2}, we investigated approximate regular Lagrangians and proved  that where the perturbation terms do not affect the kinetic energy, approximate
symmetries exist if and only if the  kinetic metric admits a nontrivial Homothetic algebra. This planted the seed that there exists a strong and deep connection between geometry and approximations. This idea was extended to Lagrangians of partial differential equations 
\cite{k4,k3,k1} and moreover a self-contained approximate Lie symmetry determining method was successfully devised in  \cite{k5}.
In the variational studies, we unveiled new higher-order approximate versions of Noether's theorem. A slightly different geometric approach can be found in  \cite{bar}. Important approximate and regular Lagrangians were considered in \cite{leach1}.
In some related work,  \cite{s1} used Noether symmetries  to select viable theories of gravity, whilst in
\cite{s2} and  \cite{s3} these   symmetries form a method to single out classical universes in quantum cosmology.
Invariant solutions emerging from Noether symmetries are discussed in \cite{s4}.  In these references,  the minisuperspaces considered are similar to the ones considered  here.

In the present work we start from the
point of view of perturbed singular Lagrangians in order to define a conditional approximate symmetry. The method essentially consists of applying a geometric approach to determine the conditional
variational symmetries. Specifically, we consider an approximate singular Lagrangian of the form:
\begin{equation}
L\left( t,x^{k},\dot{x}^{k},N,\varepsilon \right) =L_{0}\left( t,x^{k},\dot{x}%
^{k},N\right) +\varepsilon L_{1}\left( t,x^{k},\dot{x}^{k},N\right) +O\left(
\varepsilon ^{2}\right) ,  \label{ANS.01a}
\end{equation}%
where we stipulate that the exact and approximate terms are defined by the
 Lagrangians 
\begin{eqnarray}
L_{0}\left( t,x^{k},\dot{x}^{k},N\right)  &=&\frac{1}{2N}g_{ij}\dot{x}^{i}\dot{x%
}^{j}-NV_{0}\left( x^{k}\right) ,  \label{ANS.05} \\
L_{1}\left( t,x^{k},\dot{x}^{k},N\right)  &=&\frac{1}{2N}h_{ij}\dot{x}^{i}\dot{x%
}^{j}-NV_{1}\left( x^{k}\right) ,  \label{ANS.06}
\end{eqnarray}%
respectively, where the metric tensors $g_{ij}=g_{ij}\left( x^{k}\right) ,h_{ij}=h_{ij}\left(
x^{k}\right) $~, $\varepsilon$ is the perturbation parameter and dot denotes total derivative with respect to the independent
parameter $t$, i.e. $\dot{x}^{i}=\frac{dx^{i}}{dt}$.

{\bf Structure of the paper:} In \S2 we introduce the main ideas of our approach: the determining system of approximate conditional symmetries to first-order. A natural extension to nth-order Noether conditional symmetries is given in \S3. In \S4, we frame Noether's theorem in the context of our theory. Finally, 
we give explicit examples to showcase the applicability of our results and conclude in \S5 and \S6, respectively.

\section{First-Order Approximations $\left( \protect\varepsilon %
^{1}\right) $}

\label{section2}

The basic principles of our pedagogy is presented here. We point out that the nature of our geometric approach  contributes to the success of our method.
We begin with the standard first-order approximate  generator 
\begin{equation}
X=X_{0}+\varepsilon X_{1}+O\left( \varepsilon ^{2}\right), \label{ANS.G}
\end{equation}%
where we define ~\begin{equation}X_{A}=\xi _{A}\left( t,x^{k},N\right)\partial _{t}+\eta _{A}^{i}\left( t,x^{k},N\right)\partial _{i}+ \omega_A\left( t,x^{k},N\right)\p_N,\quad  A=0,1.\label{ANS.s}
\end{equation} 
We reiterate that since $N$ is considered as a degree of freedom, it appears in the same context as the $x^k$ terms.
The parameter value $A=0$ represents the exact symmetry  vector field 
while $A=1$ corresponds to the
approximate part. The generator (\ref{ANS.G}) is a Noether point
symmetry satisfying the condition%
\begin{equation}
X^{[1]}L+L\frac{d\xi }{dt}=\frac{df}{dt},
\end{equation}%
or in its approximate expanded form 
\begin{equation}
\left( X_{0}^{\left[ 1\right] }+\varepsilon X_{1}^{\left[ 1\right] }\right)
\left( L_{0}+\varepsilon L_{1}\right) +\left( L_{0}+\varepsilon L_{1}\right) 
\frac{d}{dt}\left( \xi _{0}+\varepsilon \xi _{1}\right) -\frac{d}{dt}\left(
f_{0}+\varepsilon f_{1}\right) =O\left( \varepsilon ^{2}\right) ,
\label{ANS.04}
\end{equation}%
where $f_{A}=f_{A}\left( t,x^{i},N\right) $ is the Noether boundary function. The  first prolongation is 
\begin{equation}
X_{A}^{\left[ 1\right] }=\xi _{A}\partial _{t}+\eta _{A}^{i}\partial
_{i}+\left( \dot{\eta}_{A}^{i}-\dot{x}^{i}\dot{\xi}_{A}\right) \partial _{%
\dot{x}^{i}}, \label{prolo}
\end{equation}
with $$ \dot{\eta}_{A}^{i}-\dot{x}^{i}\dot{\xi}_{A}={\eta}_{A,t}^{i}-\xi_{A,k}\dot{x}^i+{\eta}_{A,j}^{i}\dot{x}^i-\xi_{A,j}\dot{x}^i\dot{x}^j+\dot{N}{\eta}_{A,N}^{i}-\xi_{A,N}\dot{N}\dot{x}^i,$$
 minus $\p_{\dot{N}}$ in (\ref{prolo})  since the $\dot{N}$ terms  are absent in the Lagrangian.

A substitution of the Lagrange functions (\ref{ANS.05}) and (\ref{ANS.06}) into the left-hand-side of the
Noether condition (\ref{ANS.04}), we find 
\begin{eqnarray*}
X_{0}^{\left[ 1\right] }L_{0} &=&\left( \xi _{0}\partial _{t}+\eta
_{0}^{k}\partial _{k}+\left( \dot{\eta}_{0}^{k}-\dot{x}^{k}\dot{\xi}%
_{0}\right) \partial _{\dot{x}^{k}}\right) \left( \frac{1}{2N}g_{ij}\dot{x}^{i}\dot{x%
}^{j}-NV_{0}\left( x^{k}\right)\right) \\
&=&\frac{1}{2N}g_{ij,k}\eta _{0}^{k}\dot{x}^{i}\dot{x}^{j}-N\eta _{0}^{k}V_{0,k}-\omega_0\frac{1}{2N^2}g_{ij}\dot{x}^{i}\dot{x}^{j}-\omega_0V_{0}+\\
&&\frac{1}{N}\( \eta _{0,t}^{k}g_{ik}\dot{x}^{i}-\xi _{0,t}g_{ik}\dot{x}^{i}\dot{x}^{k}-\xi _{0,j}g_{ik}\dot{x}^{i}\dot{x}^{j}\dot{x}^{k}+\dot{N}\eta _{0,N}^{k}g_{ik}\dot{x}^{i}\)-\\
&&\frac{1}{N}\(\dot{N}\xi_{0,N}g_{ik}\dot{x}^{i}\dot{x}^{k}-\frac12\eta _{0,j}^{k}g_{ik}\dot{x}^{i}\dot{x}^{j}-\frac12\eta _{0,i}^{k}g_{kj}\dot{x}^{i}\dot{x}^{j} \).
\end{eqnarray*}

Similarly, 
\begin{eqnarray*}
\varepsilon X_{0}^{\left[ 1\right] }L_{1} &=&\varepsilon \Bigg(\frac{1}{2N}h_{ij,k}\eta _{0}^{k}\dot{x}^{i}\dot{x}^{j}-N\eta _{0}^{k}V_{1,k}-\omega_0\frac{1}{2N^2}h_{ij}\dot{x}^{i}\dot{x}^{j}-\omega_0V_{1}+\\
&&\frac{1}{N}\( \eta _{0,t}^{k}h_{ik}\dot{x}^{i}-\xi _{0,t}h_{ik}\dot{x}^{i}\dot{x}^{k}-\xi _{0,j}h_{ik}\dot{x}^{i}\dot{x}^{j}\dot{x}^{k}+\dot{N}\eta _{0,N}^{k}h_{ik}\dot{x}^{i}\)-\\
&&\frac{1}{N}\(\dot{N}\xi_{0,N}h_{ik}\dot{x}^{i}\dot{x}^{k}-\frac12\eta _{0,j}^{k}h_{ik}\dot{x}^{i}\dot{x}^{j}-\frac12\eta _{0,i}^{k}h_{kj}\dot{x}^{i}\dot{x}^{j} \)\Bigg), \\
\varepsilon X_{1}^{\left[ 1\right] }L_{0} &=&\varepsilon \Bigg(\frac{1}{2N}g_{ij,k}\eta _{1}^{k}\dot{x}^{i}\dot{x}^{j}-N\eta _{1}^{k}V_{0,k}-\omega_1\frac{1}{2N^2}g_{ij}\dot{x}^{i}\dot{x}^{j}-\omega_1V_{0}+\\
&&\frac{1}{N}\( \eta _{1,t}^{k}g_{ik}\dot{x}^{i}-\xi _{1,t}g_{ik}\dot{x}^{i}\dot{x}^{k}-\xi _{1,j}g_{ik}\dot{x}^{i}\dot{x}^{j}\dot{x}^{k}+\dot{N}\eta _{1,N}^{k}g_{ik}\dot{x}^{i}\)-\\
&&\frac{1}{N}\(\dot{N}\xi_{1,N}g_{ik}\dot{x}^{i}\dot{x}^{k}-\frac12\eta _{1,j}^{k}g_{ik}\dot{x}^{i}\dot{x}^{j}-\frac12\eta _{1,i}^{k}g_{kj}\dot{x}^{i}\dot{x}^{j} \)\Bigg), \\
\varepsilon ^{2}X_{1}^{\left[ 1\right] }L_{1} &=&\varepsilon ^{2}\Bigg(\frac{1}{2N}h_{ij,k}\eta _{1}^{k}\dot{x}^{i}\dot{x}^{j}-N\eta _{1}^{k}V_{1,k}-\omega_1\frac{1}{2N^2}h_{ij}\dot{x}^{i}\dot{x}^{j}-\omega_1V_{1}+\\
&&\frac{1}{N}\( \eta _{1,t}^{k}h_{ik}\dot{x}^{i}-\xi _{1,t}h_{ik}\dot{x}^{i}\dot{x}^{k}-\xi _{1,j}h_{ik}\dot{x}^{i}\dot{x}^{j}\dot{x}^{k}+\dot{N}\eta _{1,N}^{k}h_{ik}\dot{x}^{i}\)-\\
&&\frac{1}{N}\(\dot{N}\xi_{1,N}h_{ik}\dot{x}^{i}\dot{x}^{k}-\frac12\eta _{1,j}^{k}h_{ik}\dot{x}^{i}\dot{x}^{j}-\frac12\eta _{1,i}^{k}h_{kj}\dot{x}^{i}\dot{x}^{j} \)\Bigg).
\end{eqnarray*}

For the middle terms, 
\begin{equation*}
\left( L_{0}+\varepsilon L_{1}\right) \frac{d}{dt}\left( \xi
_{0}+\varepsilon \xi _{1}\right) =\dot{\xi}_{0}L_{0}+\varepsilon L_{1}\dot{%
\xi}_{0}+\varepsilon \dot{\xi}_{1}L_{0}+\varepsilon ^{2}\dot{\xi}_{1}L_{1},
\end{equation*}%
and therefore
\begin{eqnarray*}
\dot{\xi}_{0}L_{0} &=&\left( \xi _{0,t}+\xi _{0,k}\dot{x}^{k}+\dot{N}\xi _{0,N}\right) \left( 
 \frac{1}{2N}g_{ij}\dot{x}^{i}\dot{x%
}^{j}-NV_{0}\left( x^{k}\right) \right) \\
&=&\frac{1}{2N}\( \xi _{0,t} g_{ij}\dot{x}^{i}\dot{x}^{j}+ \xi _{0,k} g_{ij}\dot{x}^{i}\dot{x}^{j}\dot{x}^{k}+\dot{N} \xi _{0,N} g_{ij}\dot{x}^{i}\dot{x}^{j}\)-\\
&&\xi _{0,t}NV_0-\xi _{0,k}NV_0\dot{x}^{k}+\dot{N}N \xi _{0,N}V_0.
\end{eqnarray*}

Similarly, the rest of the expression is
\begin{eqnarray*}
\varepsilon L_{1}\dot{\xi}_{0} &=&\varepsilon \Bigg(\frac{1}{2N}\( \xi _{0,t} h_{ij}\dot{x}^{i}\dot{x}^{j}+ \xi _{0,k} h_{ij}\dot{x}^{i}\dot{x}^{j}\dot{x}^{k}+\dot{N} \xi _{0,N} h_{ij}\dot{x}^{i}\dot{x}^{j}\)-\\
&&\phantom{aaa}\xi _{0,t}NV_1-\xi _{0,k}NV_1\dot{x}^{k}+\dot{N}N \xi _{0,N}V_1\Bigg) , \\
\varepsilon \xi _{1}L_{0} &=&\varepsilon \Bigg(\frac{1}{2N}\( \xi _{1,t} g_{ij}\dot{x}^{i}\dot{x}^{j}+ \xi _{1,k} g_{ij}\dot{x}^{i}\dot{x}^{j}\dot{x}^{k}+\dot{N} \xi _{1,N} g_{ij}\dot{x}^{i}\dot{x}^{j}\)-\\
&&\phantom{aaa}\xi _{1,t}NV_0-\xi _{1,k}NV_0\dot{x}^{k}+\dot{N}N \xi _{1,N}V_0\Bigg) , \\
\varepsilon ^{2}\xi _{1}L_{1} &=&\varepsilon^2 \Bigg(\frac{1}{2N}\( \xi _{1,t} h_{ij}\dot{x}^{i}\dot{x}^{j}+ \xi _{1,k} h_{ij}\dot{x}^{i}\dot{x}^{j}\dot{x}^{k}+\dot{N} \xi _{1,N} h_{ij}\dot{x}^{i}\dot{x}^{j}\)-\\
&&\phantom{aaa}\xi _{1,t}NV_1-\xi _{1,k}NV_1\dot{x}^{k}+\dot{N}N \xi _{1,N}V_1\Bigg) .
\end{eqnarray*}

The boundary terms produce the expression 
\begin{equation*}
\dot{f}_{0}+\varepsilon \dot{f}_{1}=\left( f_{0,t}+f_{0,k}\dot{x}^{k}+\dot{N}f_{0,N}\right)
+\varepsilon \left( f_{1,t}+f_{1,k}\dot{x}^{k}+\dot{N}f_{1,N}\right) .
\end{equation*}

Next, we distinguish terms based on the order of $\varepsilon $, viz. we obtain the equation for
$\varepsilon ^{0}:$
\begin{eqnarray*}
&&\frac{1}{2N}g_{ij,k}\eta _{0}^{k}\dot{x}^{i}\dot{x}^{j}-N\eta _{0}^{k}V_{0,k}-\omega_0\frac{1}{2N^2}g_{ij}\dot{x}^{i}\dot{x}^{j}-\omega_0V_{0}+\\
&&\frac{1}{N}\( \eta _{0,t}^{k}g_{ik}\dot{x}^{i}-\xi _{0,t}g_{ik}\dot{x}^{i}\dot{x}^{k}-\xi _{0,j}g_{ik}\dot{x}^{i}\dot{x}^{j}\dot{x}^{k}+\dot{N}\eta _{0,N}^{k}g_{ik}\dot{x}^{i}\)-\\
&&\frac{1}{N}\(\dot{N}\xi_{0,N}g_{ik}\dot{x}^{i}\dot{x}^{k}-\frac12\eta _{0,j}^{k}g_{ik}\dot{x}^{i}\dot{x}^{j}-\frac12\eta _{0,i}^{k}g_{kj}\dot{x}^{i}\dot{x}^{j} \)+\\
&&\frac{1}{2N}\( \xi _{0,t} g_{ij}\dot{x}^{i}\dot{x}^{j}+ \xi _{0,k} g_{ij}\dot{x}^{i}\dot{x}^{j}\dot{x}^{k}+\dot{N} \xi _{0,N} g_{ij}\dot{x}^{i}\dot{x}^{j}\)-\\
&&\xi _{0,t}NV_0-\xi _{0,k}NV_0\dot{x}^{k}+\dot{N}N \xi _{0,N}V_0\\
&&=f_{0,t}+f_{0,k}\dot{x}^{k}+\dot{N}f_{0,N}.
\end{eqnarray*}
It is now necessary to separate monomials to obtain  the determining system of
equations:
\begin{eqnarray}
\(\dot{N}\dot{x}^i\dot{x}^j\) &:&\nonumber\\
&&-\frac{1}{2N}\xi_{0,N}g_{ij}=0,\label{s1}\\
\(\dot{x}^i\dot{x}^j\dot{x}^k\) &:&\nonumber\\
&&-\frac{1}{2N}\xi_{0,k}g_{ik}=0,\label{s2}\\
\(\dot{N}\dot{x}^i\) &:&\nonumber\\
&&\frac{1}{N}\eta^k_{0,N}g_{ik}=0,\label{s3}\\
\(\dot{x}^i\dot{x}^j\) &:&\nonumber\\
&&{\cal L}_{\eta_0} g_{ik}=\(\eta^k_{0}\(\ln{V_0}\)_{,k}-\frac{f_{0,t}}{NV_0}\)g_{ik},\label{s4}\\
\(\dot{x}^k\) &:&\nonumber\\
&&-f_{0,k}-\xi_{0,k}NV_0+\frac1N \eta_{0,t}^kg_{ik}=0,\label{s5}\\
\(\dot{N}\) &:&\nonumber\\
&&-f_{0,N}+N\xi_{0,N}V_0=0,\label{s6}\\
\(1\) &:&\nonumber\\
&&\omega_0=-N\eta^k_{0}\(\ln{V_0}\)_{,k}-N\xi_{0,t}-\frac{f_{0,t}}{V_0},\quad V_0\ne0.\label{s7}
\end{eqnarray}
The $L_{\eta _{j}}$ refers to the geometric derivative or Lie
derivative operator along $\eta _{j}$.
From Eq. (\ref{s1})-(\ref{s7}), we have that $f_0=f_0\(t,x^i\), \eta^k_0=\eta_0^k\(t,x^k\)$ and $\xi_0=\xi_0\(t\)$.

For the determining equation involving 
$\varepsilon ^{1},$ we find:  
\begin{eqnarray*}
&&\frac{1}{2N}h_{ij,k}\eta _{0}^{k}\dot{x}^{i}\dot{x}^{j}-N\eta _{0}^{k}V_{1,k}-\omega_0\frac{1}{2N^2}h_{ij}\dot{x}^{i}\dot{x}^{j}-\omega_0V_{1}+\\
&&\frac{1}{N}\( \eta _{0,t}^{k}h_{ik}\dot{x}^{i}-\xi _{0,t}h_{ik}\dot{x}^{i}\dot{x}^{k}-\xi _{0,j}h_{ik}\dot{x}^{i}\dot{x}^{j}\dot{x}^{k}+\dot{N}\eta _{0,N}^{k}h_{ik}\dot{x}^{i}\)-\\
&&\frac{1}{N}\(\dot{N}\xi_{0,N}h_{ik}\dot{x}^{i}\dot{x}^{k}-\frac12\eta _{0,j}^{k}h_{ik}\dot{x}^{i}\dot{x}^{j}-\frac12\eta _{0,i}^{k}h_{kj}\dot{x}^{i}\dot{x}^{j} \)+ \\
&&\frac{1}{2N}g_{ij,k}\eta _{1}^{k}\dot{x}^{i}\dot{x}^{j}-N\eta _{1}^{k}V_{0,k}-\omega_1\frac{1}{2N^2}g_{ij}\dot{x}^{i}\dot{x}^{j}-\omega_1V_{0}+\\
&&\frac{1}{N}\( \eta _{1,t}^{k}g_{ik}\dot{x}^{i}-\xi _{1,t}g_{ik}\dot{x}^{i}\dot{x}^{k}-\xi _{1,j}g_{ik}\dot{x}^{i}\dot{x}^{j}\dot{x}^{k}+\dot{N}\eta _{1,N}^{k}g_{ik}\dot{x}^{i}\)-\\
&&\frac{1}{N}\(\dot{N}\xi_{1,N}g_{ik}\dot{x}^{i}\dot{x}^{k}-\frac12\eta _{1,j}^{k}g_{ik}\dot{x}^{i}\dot{x}^{j}-\frac12\eta _{1,i}^{k}g_{kj}\dot{x}^{i}\dot{x}^{j} \) +\\
&&\frac{1}{2N}\( \xi _{0,t} h_{ij}\dot{x}^{i}\dot{x}^{j}+ \xi _{0,k} h_{ij}\dot{x}^{i}\dot{x}^{j}\dot{x}^{k}+\dot{N} \xi _{0,N} h_{ij}\dot{x}^{i}\dot{x}^{j}\)-\\
&&\phantom{aaa}\xi _{0,t}NV_1-\xi _{0,k}NV_1\dot{x}^{k}+\dot{N}N \xi _{0,N}V_1+ \\
&&\frac{1}{2N}\( \xi _{1,t} g_{ij}\dot{x}^{i}\dot{x}^{j}+ \xi _{1,k} g_{ij}\dot{x}^{i}\dot{x}^{j}\dot{x}^{k}+\dot{N} \xi _{1,N} g_{ij}\dot{x}^{i}\dot{x}^{j}\)-\\
&&\phantom{aaa}\xi _{1,t}NV_0-\xi _{1,k}NV_0\dot{x}^{k}+\dot{N}N \xi _{1,N}V_0= \\
&& f_{1,t}+f_{1,k}\dot{x}^{k}+\dot{N}f_{1,N}.
\end{eqnarray*}

Note that we impose the restriction that terms involving $\varepsilon ^{2}$
vanish.
Separating coefficients here leads to the determining system: 
\begin{eqnarray}
\(\dot{N}\dot{x}^i\dot{x}^j\) &:&\nonumber\\
&&-\frac{1}{2N}\xi_{1,N}g_{ij}=0,\label{s11}\\
\(\dot{x}^i\dot{x}^j\dot{x}^k\) &:&\nonumber\\
&&-\frac{1}{2N}\xi_{1,k}g_{ik}=0,\label{s21}\\
\(\dot{N}\dot{x}^i\) &:&\nonumber\\
&&\frac{1}{N}\eta^k_{1,N}g_{ik}=0,\label{s31}\\
\(\dot{x}^i\dot{x}^j\) &:&\nonumber\\
&&{\cal L}_{\eta_0} h_{ik}+{\cal L}_{\eta_1} g_{ik}=\({\xi_{0,t}}+\frac{w_{0}}{N}\)h_{ik}+\({\xi_{1,t}}+\frac{w_{1}}{N}\)g_{ik},\label{s41}\\
\(\dot{x}^k\) &:&\nonumber\\
&&-f_{1,k}-\xi_{0,k}NV_1-\xi_{1,k}NV_0+\frac1N \eta_{0,t}^kh_{ik}+\frac1N \eta_{1,t}^kg_{ik}=0,\label{s51}\\
\(\dot{N}\) &:&\nonumber\\
&&-f_{1,N}+N\xi_{0,N}V_1+N\xi_{1,N}V_0=0,\label{s61}\\
\(1\) &:&\nonumber\\
&&\frac{\omega_0}{V_0}+\frac{\omega_1}{V_1}=-N\eta^k_{0}\frac{\(\ln{V_1}\)_{,k}}{V_0}-\frac{N\xi_{0,t}}{V_0}-N\eta^k_{1}\frac{\(\ln{V_0}\)_{,k}}{V_1}-\frac{N\xi_{1,t}}{V_1}\nonumber\\
&&\phantom{aaaaa}-\frac{f_{1,t}}{V_0V_1},\quad V_{0-1}\ne0.\label{s71}
\end{eqnarray}

Also $f_1=f_1\(t,x^i\), \eta^k_1=\eta_1^k\(t,x^k\)$ and $\xi_1=\xi_1\(t\)$.
Finally, the system of equations (\ref{s1})-(\ref{s7}) and (\ref{s11})-(\ref%
{s71}) provide the approximate Noether symmetry conditions for the perturbed
Lagrangian (\ref{ANS.01a}) defined by equations (\ref{ANS.05}), (\ref%
{ANS.06}) and in correspondence with the Noether symmetry vector (\ref{ANS.G}). In order to enhance the applicability of these geometric conditions, we now derive the  higher-order version of this approach.

\section{Generalizations to $\left( \protect\varepsilon %
^{n}\right) $.}

Let us extend our analysis to  approximations 
of any order, that is  $\varepsilon ^{n}.$ 
To initailize this generalization, we consider the approximate symmetry generator%
\begin{equation}
X=X_{0}+\dsum\limits_{\gamma =1}^{n}\varepsilon ^{\gamma }X_{\gamma
}+O\left( \varepsilon ^{\gamma +1}\right).\label{G00}
\end{equation}%

Analogous to  the Noether condition (\ref{ANS.04}), we impose the   the generalized  condition 
\begin{eqnarray}
&&\left( X_{0}^{\left[ 1\right] }+\dsum\limits_{\gamma =1}^{n}\varepsilon
^{\gamma }X_{\gamma }^{\left[ 1\right] }\right) \left( L_{0}+\varepsilon
L_{1}\right) +\left( L_{0}+\varepsilon L_{1}\right) \frac{d}{dt}\left( \xi
_{0}+\dsum\limits_{\gamma =1}^{n}\varepsilon ^{\gamma }\xi _{\gamma }\right)
\notag \\
&&-\frac{d}{dt}\left( f_{0}+\dsum\limits_{\gamma =1}^{n}\varepsilon ^{\gamma
}f_{\gamma }\right) =O\left( \varepsilon ^{\gamma +1}\right) .
\end{eqnarray}%
In lieu of these generalizations, we state the following determining system for higher-order approximate Noether symmetries. 
The derivation of this system follows the same procedure outlined in  Section 2,  but for the economy of space we simply state the relevant formulae. To this
end, the Noether symmetry conditions for $\left( \varepsilon \right) ^{0}$ are the same as before, namely Eqs. (\ref{s1})-(\ref{s7}). On the other hand for 
$\left( \varepsilon \right) ^{\gamma }~,~\gamma =1\ldots n:$  
we obtain the determining system
\begin{eqnarray}
\(\dot{N}\dot{x}^i\dot{x}^j\) &:&\nonumber\\
&&-\frac{1}{2N}\xi_{\gamma,N}g_{ij}=0,\label{s12}\\
\(\dot{x}^i\dot{x}^j\dot{x}^k\) &:&\nonumber\\
&&-\frac{1}{2N}\xi_{\gamma,k}g_{ik}=0,\label{s22}\\
\(\dot{N}\dot{x}^i\) &:&\nonumber\\
&&\frac{1}{N}\eta^k_{\gamma,N}g_{ik}=0,\label{s32}\\
\(\dot{x}^i\dot{x}^j\) &:&\nonumber\\
&&{\cal L}_{\eta_{\gamma-1}} h_{ik}+{\cal L}_{\eta_\gamma} g_{ik}=\({\xi_{\gamma-1,t}}+\frac{w_{\gamma-1}}{N}\)h_{ik}+\({\xi_{\gamma,t}}+\frac{w_{\gamma}}{N}\)g_{ik},\label{s42}\\
\(\dot{x}^k\) &:&\nonumber\\
&&-f_{\gamma,k}-\xi_{\gamma-1,k}NV_1-\xi_{\gamma,k}NV_0+\frac1N \eta_{\gamma-1,t}^kh_{ik}+\frac1N \eta_{\gamma,t}^kg_{ik}=0,\label{s52}\\
\(\dot{N}\) &:&\nonumber\\
&&-f_{\gamma,N}+N\xi_{\gamma-1,N}V_1+N\xi_{\gamma,N}V_0=0,\label{s62}\\
\(1\) &:&\nonumber\\
&&\frac{\omega_\gamma}{V_0}+\frac{\omega_{\gamma-1}}{V_1}=-N\eta^k_{\gamma-1}\frac{\(\ln{V_1}\)_{,k}}{V_0}-\frac{N\xi_{\gamma-1,t}}{V_0}-N\eta^k_{\gamma}\frac{\(\ln{V_0}\)_{,k}}{V_1}-\frac{N\xi_{\gamma,t}}{V_1}\nonumber\\
&&\phantom{aaaaaaaaaaaa}-\frac{f_{\gamma,t}}{V_0V_1},\quad V_{0-1}\ne0.\label{s72}
\end{eqnarray}

The importance of this system is that its solution provides approximate and conditional symmetries at higher-order perturbations.

\section{Noether Integrals}

By Noether's theorem \cite{noet} the
 symmetry vector field (\ref{ANS.s}) with $A=0$~ for the Lagrangian  (\ref{ANS.05})  with 
boundary term $f$, admits the conservation law:%
\begin{eqnarray*}
I_0&=&\xi_0 \( \dot{x}^{i}\frac{\partial L_0}{\partial \dot{x}^{i}}+ \dot{N}\frac{\partial L_0}{\partial \dot{N}}-L_0\)-\frac{\partial L_0}{\partial \dot{x}^{i}}\eta_0 ^{i}-\frac{\partial L_0}{\partial \dot{N}}w_0+f_0,\\
&=& \xi_0 \(\frac{1}{2N}g_{ij}\dot{x}^{i}\dot{x%
}^{j}+NV_{0}\)-\frac{1}{2N}g_{ik}\dot{x%
}^{i}\eta_0 ^{i}+f_0\\
&=& \xi_0 H_{0}-\frac{1}{2N}g_{ij}\dot{x%
}^{j}\eta_0 ^{i}+f_0,
\end{eqnarray*}%
since there are no $\dot{N}$ terms in the Lagrangian and where $H_0$ is the exact Hamiltonian.

In a similar way, for any approximate Lagrangian  (\ref{ANS.01a})  and an approximate Noether generator  (\ref{ANS.G}),  we derive the first-order approximate
part, $I_{1}$ as follows:%
\begin{eqnarray}
I_{1} &=& H_{0}\xi _{1}-\frac{1}{2N}g_{ij}\dot{x%
}^{j}
\eta _{1}^{i}+f_{1} +\xi _{0}H_{1}-\frac{1}{2N}h_{ij}\dot{x%
}^{j}\eta _{0}^{i}.  \label{cone2}
\end{eqnarray}

A generalization of this idea to the higher-order case of $\varepsilon $,
with the symmetry generator (\ref{G00})
leads us to deduce the formulae for the associated Noether integrals, viz. 
\begin{eqnarray}
I_{\gamma } &=& H_{0}\xi _{\gamma }-\frac{1}{2N}g_{ij}\dot{x%
}^{j}\eta _{\gamma }^{i}+f_{\gamma } +\xi _{\gamma -1}H_{1}-%
\frac{1}{2N}h_{ij}\dot{x%
}^{j}\eta _{\gamma -1}^{i}.
\end{eqnarray}

The function $I_{0}$ or $I_{\gamma}$ gives rise to a so called  `weak' conservation law in the sense that one needs to impose the 
constraint condition
$\frac{\p L}{\p N}$=0 in order for $D_t{I}=0$; we illustrate this in the next section.

\section{Applications}

\label{section4}

Now that we have  developed  an explicit method of deriving the conditional and
approximate Noether symmetry conditions, we may tackle some examples. The progression from the determining system to the symmetry generator is as follows. With the help of Eqs. (\ref{s1})-(\ref{s7}) and Eqs. (\ref{s12})-(\ref{s72}) we will
be led to the conditions that the Noether symmetries must satisfy. The solution of the set of these conditions must be done sequentially in order to acquire the conditional symmetry generators.  This process is 
straightforward, albeit lengthy and so we merely
 list the pertinent  results. 
That is, we present the approximate Noether symmetries and first integrals for each example. 

\subsection{Case A:}
Consider the Lagrange functions
$$L_0=\frac{1}{2N}\dot{x}^2-\frac{N}{2}x^2,~L_1=- NV_1 \frac{x^3}{3},$$
with corresponding Hamiltonian functions
$$H_0=\frac{1}{2N}\dot{x}^2+\frac{N}{2}x^2,~H_1=NV_1 \frac{x^3}{3}.$$
Since we would like to compare the symmetries obtained under fixing the lapse function with allowing the lapse to be a degree of freedom, for this first case we present both results. As the reader shall see, the results differ substantially. 
\begin{itemize}
\item  For the exact and approximate Noether symmetries under constant lapse $N=1$, the approximate symmetries are \cite{gov}, 
$$\begin{array}{lc}
&Y^1=\partial_t,\quad f=0, \\
&Y^2=\sin(2 t)\partial_\phi+ \cos(2 t)x\partial_x,\quad f=-x^2\sin(2t), \\
&Y^3=\cos(2 t)\partial_\phi- \sin(2t)x\partial_x,\quad f=-x^2\cos(2t), \\
&Y^4=\sin(t)\partial_x,\quad f=x\cos(t), \\
&Y^5=\cos(t)\partial_x,\quad f=-x\sin(t),\\
&Y^6=\sin(t)\partial_x+\varepsilon\(\frac43 V_1 \cos(t)\p_t-\frac23V_1\sin(t)\p_x\),\quad f=x\cos(t)-\varepsilon\(\frac{V_1x^2}{3}\cos(t)\) \\
&Y^7=\cos(t)\partial_x+\varepsilon\(-\frac43 V_1 \sin(t)\p_t-\frac23V_1\cos(t)\p_x\),\quad f=-x\sin(t)+\varepsilon\(\frac{V_1x^2}{3}\sin(t)\).
\end{array}$$
\item For the conditional Noether symmetries with $N=N(t)$ we obtain
$$X^{Ai}=T(t)~\p_t+\frac{1}{x}\p_x-N\(\frac{d }{d t}T(t)+\frac{2}{x^2}\)\p_N, \quad f=0,$$
$$X^{Aii}=T(t)~\p_t+\frac{1}{x}\p_x-N\(\frac{d }{d t}T(t)+\frac{2}{x^2}\)\p_N+\varepsilon\(T(t)~\p_t-\frac{V_1}{3}\p_x-N\frac{d }{d t}T(t)~\p_N\),~ f=0.$$
\item For the approximate Noether integrals we have 
\begin{eqnarray*}
I(X^{Ai})&=&T(t)H_0-\frac1x\frac{\dot{x}}{N},\\
I(X^{Aii})&=&T(t)H_0-\frac1x\frac{\dot{x}}{N}+\varepsilon\(T(t)H_0+T(t)H_1+\frac{V_1}{3}\frac{\dot{x}}{N}\).
\end{eqnarray*}
\end{itemize}
With regard to the conservation laws corresponding to conditional symmetry vectors, as an example: $$D_t I\(X^{Ai}\)=\frac{2}{x^2}\frac{\p L}{\p N},$$ is a
multiple of the constraint equation, rather than strictly zero. 

\subsection{Case B:}
In this case we take the Lagrangians
$$L_0=\frac{1}{2N}6x\dot{x}^2+2\Lambda {N}x^3,~L_1= N \frac{V_1}{x^2},$$
with Hamiltonians
$$H_0=\frac{1}{2N}6x\dot{x}^2-2\Lambda {N}x^3,~H_1= -N \frac{V_1}{x^2}.$$
\begin{itemize}
\item For the exact and approximate Noether symmetries we find 
$$X^{Bi}=T(t)~\p_t+\frac{1}{x^2}\p_x-N\(\frac{d }{d t}T(t)+\frac{3}{x^3}\)\p_N, \quad f=0,$$
$$X^{Bii}=T(t)~\p_t+\frac{1}{x^2}\p_x-N\(\frac{d }{d t}T(t)+\frac{3}{x^3}\)\p_N+\varepsilon\(\frac{5}{32}\frac{V_1}{x^7 \Lambda}\p_x+N\frac{65~V_1}{32~\Lambda~x^8}~\p_N\),~ f=0,$$
$$X^{Biii}=\varepsilon\(t\p_t+x\ln x\p_x-3N\(\ln x +\frac13\)\p_N\),\quad f=0,$$
$$X^{Biv}=\varepsilon\p_t,\quad f=0,\quad X^{Bv}=\varepsilon \(x\p_x-3N\p_N\),\quad f=0.$$
\item The approximate Noether integrals are 
\begin{eqnarray*}
I(X^{Bi})&=&T(t)H_0-\frac{1}{x^2}\frac{x\dot{x}}{N},\\
I(X^{Bii})&=&T(t)H_0-\frac{1}{x^2}\frac{x\dot{x}}{N}-\varepsilon\(\frac{30V_1\dot{x}}{32\Lambda N x^6}\),\\
I(X^{Biii})&=&\varepsilon\(tH_0-\frac{6x^2\ln x~\dot{x}}{N}\),\\
I(X^{Biv})&=&\varepsilon H_0,\\
I(X^{Bv})&=&-\varepsilon\frac{6x^2\dot{x}}{N}.
\end{eqnarray*}
\end{itemize}

\subsection{Case C:}
Suppose we have the Lagrangians
$$L_0=\frac{1}{2N}\dot{x}^2-{N}{V_0^2}x^2,~L_1=- NV_1 \frac{\exp{(x)}}{2},$$
and Hamiltonian functions
$$H_0=\frac{1}{2N}\dot{x}^2+{N}{V_0^2}x^2,~H_1= NV_1 \frac{\exp{(x)}}{2}.$$
\begin{itemize}
\item The Noether symmetries are 
\begin{eqnarray*}
X^{Ci}&=&T(t)~\p_t+\frac{1}{x}\p_x-N\(\frac{d }{d t}T(t)+\frac{2}{x^2}\)\p_N, \quad f=0,\\
X^{Cii}&=&T(t)~\p_t+\frac{1}{x}\p_x-N\(\frac{d }{d t}T(t)+\frac{2}{x^2}\)\p_N+\\
&&\varepsilon\(T(t)~\p_t-\frac14\frac{V_1\exp^x}{V_0^2x^3}\p_x-N\(\frac{d }{d t}T(t)+\frac12 \frac{V_1\(x-3\)\exp^x}{V_0^2x^4}\)~\p_N\),\\&&\phantom{aaaaaaaaaaaaaaaaa} \quad f=0.
\end{eqnarray*}
\item For the approximate Noether integrals we have 
\begin{eqnarray*}
I(X^{Ci})&=&T(t)H_0-\frac1x\frac{\dot{x}}{N},\\
I(X^{Cii})&=&T(t)H_0-\frac1x\frac{\dot{x}}{N}+\varepsilon\(T(t)H_0+T(t)H_1-\frac14\frac{V_1\exp^x}{V_0^2x^3}\frac{\dot{x}}{N}\).
\end{eqnarray*}
\end{itemize}

\subsection{Case D:}
In the last case, let us consider
$$L_0=\frac{1}{2N}\dot{x}^2+N\frac{x^3}{3},~L_1=N \frac{V_1 x^n}{n},\quad n\ne0,3$$ and
$$H_0=\frac{1}{2N}\dot{x}^2-N\frac{x^3}{3},~H_1=-N \frac{V_1 x^n}{n}.$$
\begin{itemize}
\item Now, the Noether symmetries are found to be
\begin{eqnarray*}
X^{Di}&=&T(t)~\p_t+\frac{1}{x^{3/2}}\p_x-N\(\frac{d }{d t}T(t)+\frac{3}{x^{5/2}}\)\p_N, \quad f=0,\\
X^{Dii}&=&T(t)~\p_t+\frac{1}{x^{3/2}}\p_x-N\(\frac{d }{d t}T(t)+\frac{3}{x^{5/2}}\)\p_N\\
&&
+\varepsilon\Bigg(T(t)~\p_t+ \frac32\,{\frac {V_1 \, \left( n+3 \right) {x_{{}}}^{n-3}}{{x_{{}}}^{
3/2} \left( n-3 \right) n}}
\p_x\\&&-N{\frac {\Big( \frac92\,{x_{{}}}^{3}V_1 \, \left( n+3
 \right) {x_{{}}}^{n-3}+ \big(  \frac{d }{d t}T(t) ~{x_{{}}}^{11/2}n-3\,V_1 \,
 \left( n+3 \right) {x_{{}}}^{n} \big)~  \left( n-3 \right)  \Big) 
}{{x_{{}}}^{11/2} \left( n-3 \right) n}}
~\p_N\Bigg),\\&&\phantom{aaaaaaaaaaaaaaaaa} \quad f=0.
\end{eqnarray*}
\item The corresponding Noether integrals are
\begin{eqnarray*}
I(X^{Di})&=&T(t)H_0-\frac{1}{x^3/2}\frac{\dot{x}}{N},\\
I(X^{Dii})&=&T(t)H_0-\frac{1}{x^3/2}\frac{\dot{x}}{N}+\varepsilon\(T(t)H_0+T(t)H_1-\frac32\,{\frac {V_1 \, \left( n+3 \right) {x_{{}}}^{n-3}}{{x_{{}}}^{
3/2} \left( n-3 \right) n}}\frac{\dot{x}}{N}\).
\end{eqnarray*}
\end{itemize}

\section{Conclusion and Outlook}

\label{section5}

In this work, we bypassed the usual procedure of gauge fixing the lapse function  to obtain a constrained and approximate action quadratic in velocities. This combination created a challenging problem from a symmetry perspective, especially in the presence of a broader space of variables and/or increasingly higher-order perturbations. The addition of a geometric approach allowed us to
examine the fate of the resultant Noether symmetries. We encountered
the coefficient of $\p_t$ as an unrestricted function of time, both exactly and approximately, a special feature of singular Lagrangians owing to the time reparametrisation invariance. This coincides with the results found in  \cite{p2} for the exact case. Lastly, this study showed that  constraint dependent  variational symmetries are obtainable in an approximate sense.

\bigskip

\textbf{Acknowledgments:} We acknowledge the
financial support from the National Research Foundation of South Africa
(99279). 

\appendix%



\end{document}